\documentclass[preprint,showpacs,nofootinbib,prd,aps]{revtex4-1}
\usepackage{graphicx,amstext,amssymb}
\usepackage[usenames]{color}
\begin{document}
\title{Thermalization, evolution, and LHC observables in an integrated hydrokinetic model of \emph{A+A} collisions}

\author{V.Yu. Naboka$^1$}
\author{Iu.A. Karpenko$^{1,2}$}
\author{Yu.M. Sinyukov$^{1}$}
\affiliation{$^1$Bogolyubov Institute for Theoretical Physics, Metrolohichna str. 14b, 03680 Kiev}

\affiliation{$^2$ Frankfurt Institute for Advanced Studies, Ruth-Moufang-Stra\ss e 1, 60438 Frankfurt am Main}

\begin{abstract}
A further development of the evolutionary picture of \emph{A+A} collisions, which we
call the integrated HydroKinetic Model (iHKM), is proposed. The model comprises a generator of the
initial state GLISSANDO, pre-thermal dynamics of \emph{A+A} collisions leading to
thermalization, subsequent relativistic viscous hydrodynamic expansion of
quark-gluon and hadron medium (vHLLE), its particlization, and finally
hadronic cascade ultrarelativistic QMD. We calculate mid-rapidity charged-particle multiplicities, pion, kaon, and
antiproton spectra, charged-particle elliptic flows, and pion interferometry radii for Pb+Pb collisions at the energies available at the CERN Large Hadron Collider, $\sqrt{s} = 2.76$ TeV, at different centralities. We find that the best description
of the experimental data is reached when the initial states are attributed to the very small initial
time 0.1 fm/c,  the pre-thermal stage (thermalization process)   lasts at least until  1 fm/c, and the shear viscosity at the hydrodynamic stage of the matter evolution has its minimal value, $\eta/s = \frac{1}{4\pi}$. At the same time it is observed  that the various momentum anisotropies of the initial states, different initial and relaxation times, as  well as even a treatment of the pre-thermal stage within just viscous or ideal hydrodynamic approach,
leads sometimes to worse but nevertheless similar results, {\it if}  the normalization of maximal initial energy density in most
central events is adjusted to reproduce the final hadron multiplicity in each scenario.
This can explain a good enough data description in numerous variants of
hybrid models without a prethermal stage when the initial energy densities are
defined  up to a common factor.
\end{abstract}


 \maketitle

 \section{Introduction}

Hydrodynamics is considered  now as the basic  part of a spatiotemporal
picture of the matter evolution in the processes of ultrarelativistic heavy ion collisions (see recent reviews \cite{Kodama,Song}). To complete the
description of \emph{A+A} collision processes, hydrodynamics must
be supplied with a generator of an initial non-equilibrated state, pre-thermal dynamics which forms the initial near locally equilibrated conditions for hydro-evolution, and prescription for particle production during the breakup of the
continuous medium at the final stage of the matter expansion.

As for the initial state, since it fluctuates on an event-by-event basis, Monte Carlo event generators
are widely used to simulate it in relativistic
$A+A$ collisions. The most commonly used models of initial state
are the MC-Glauber (Monte Carlo Glauber) \cite{MCG}, MC-KLN (Monte Carlo
Kharzeev-Levin-Nardi) \cite{MC-KLN}, EPOS (parton-based Gribov-Regge model) \cite{EPOS}, EKRT (perturbative QCD + saturation model) \cite{EKRT, EKRT-2}, and IP-Glasma (impact parameter dependent
glasma) \cite{IPG}.\footnote{Note that the last two models are able to
reproduce correctly the centrality systematics of the $\delta v_2$ \cite{EKRT-2,IPG-2}.
} The last model also includes non-trivial non-equilibrium
dynamics of the gluon fields which, however, does not lead to a proper
equilibration. To apply these models to data description some
thermalization process has to be assumed. Evidently, in order to reduce
uncertainties of results obtained by means of hydrodynamical models,
one needs to convert a far-from-equilibrium initial state of matter in a
nucleus-nucleus collision to a close to locally equilibrated one by means
of a reasonable pre-equilibrium dynamics. A relaxation time method \cite{preth}, initially developed for the post-hydrodynamic stage, is modified for applications to pre-thermal dynamics.

As for the post-hydrodynamic (afterburner) stage, good progress was made in understanding and
modeling the breakup of hadron matter into hadron gas at the late stage of expansion. Typically, particlization, i.e., the transition from the hadron (or quark-gluon) fluid
to hadronic gas expansions,  is described by means of the so-called Cooper-Frye prescription within hybrid
models. In such an approach, the sudden conversion of the fluid to particles happens
 at a hypersurface of hadronization or chemical
freeze-out. It has long been known that such a
matching prescription has problems with the energy-momentum
conservation laws when fluid is converted to particles at the
hypersurface which contains non-space-like parts (see up-to-date review \cite{Kodama} and references therein). These problems can
be avoided by using the HydroKinetic Model (HKM) that was proposed in Ref.
\cite{hydrokin-1} and further developed in Refs. \cite{hydrokin-2, hydrokin-2a}
(see also Ref. \cite{hydrokin-3}). In basic HKM, particlization is considered not as a sudden, as in hybrid models, but as a realistic continuous process.\footnote{For recent
discussions of the particlization procedure see, e.g., Refs.
\cite{hydrokin-3,convert}).} In such an approach the description of continuous particle emission from the fluid  is based not on the distribution function  but on the so-called
escape function \cite{hydrokin-1, hydrokin-3}.

In Refs. \cite{preth, Naboka} the method of escape probabilities, similar to that used in HKM formalism for post-thermal dynamics, is applied to the pre-thermal one. It makes it possible to describe the pre-thermal evolution of the energy-momentum tensor using three free parameters: the time of the initial (non-equilibrium) state formation, mean relaxation time, and time of thermalization, say, 1 fm/c. This phenomenological model allows one to provide  initialization of the  hydrodynamical evolution, and is a more general and simple alternative to the so-called ``anisotropic hydrodynamics'' \cite{RSF, aniz, Mart-1, Mart-2, Mart-3}, as discussed in detail in \cite{Naboka}.

The target energy-momentum tensor of the pre-thermal dynamics of \emph{A+A} collisions, which the initially created  system reaches during evolution, is the tensor of viscous relativistic hydrodynamics. The further evolution is governed by the equations of the Israel-Stewart framework of relativistic viscous hydrodynamics until the lowest possible temperature when the system is still close to local thermal and chemical equilibrium. It defines the hypersurface which is often called the chemical freeze-out isotherm.
The particlization procedure starts typically at this very hypersurface. The hypersurface contains non-space-like parts that formally enforce one to use the HKM continuous particlization procedure, to eliminate the problem with violation of the energy-momentum conservation. However, with almost smooth (averaged) initial conditions, the effects originated from these parts are quite small and do not affect significantly  the observables  \cite{hydrokin-2a}.\footnote{The reason is that at small proper times $\tau$ the three-dimensional areas of non-space-like parts are small, and for large times the hydro-velocities at the corresponding parts are large.} Moreover, in viscous hydrodynamics the effects of violation of local equilibrium, which is produced in HKM for an ideal fluid, are partially taken into account. Therefore, summarizing all that and accounting for very time consuming calculations in HKM, we will use in this paper the sudden transition from hadron matter to hadron gas, similar to the hybrid models. In future studies we plan to include the option of continuous HKM particlization for situations when undesirable  contributions from non-space-like parts are fairly large: it can be for  $4\pi$- geometry observables and event-by-event analysis. Considering also that an important component of the full model is the pre-thermal evolution where the description is built on the same idea as the basic HKM formalism, we will call the complete model, which includes generation of the initial state, its thermalization, sudden (with the possible option of ``continuous'') particlization, subsequent ultrarelativistic QMD (UrQMD) hadronic  cascade \cite{urqmd}, and one- and multi- particle spectra formation, the integrated HydroKinetic Model (iHKM).

The paper is organized as follows. Section 2 includes subsections that describe the ingredients of the complete iHKM. Section 3 is devoted to the results and discussions, while the Section 4 summarizes the physical meaning of obtained results.

\section{The model description}
As discussed in the Introduction the integrated HydroKinetic Model (iHKM) consists of five ingredients which we describe below.

\subsection{Initial state}
Though prethermal dynamics is used to initialize hydrodynamics, it itself needs the initial state of the matter at the starting time $\tau_0$, say $\tau_0=0.1$ or 0.5 fm/c, when one can speak about energy density distribution in non equilibrated matter. Following Ref. \cite{{Krasnitz}}, an appropriate scale
controlling the formation of gluons with a physically well-defined energy is roughly $3/g^2\mu$, where $g^2 = 4\pi\alpha_s$ and $\mu^2$ are dimensionless parameters, which
is the variance of the Gaussian weight over the color charges
$\rho$ of partons. The estimate of this time gives $\tau_0=0.1$ fm/c for the top energy available at the BNL Relativistic heavy Ion Collider (RHIC) and $\tau_0=0.07$ fm/c for the CERN Large Hadron Collider (LHC) full energy \cite{Lappi}. We will utilize  $\tau_0=0.1$ fm/c as the default value for the LHC energy $\sqrt{s_{NN}}= 2.76$ TeV; however, we will use also $\tau_0=0.5$ fm/c to analyze the corresponding variation of the results.

We attribute the initial state to the time $\tau_0$. The generation of the initial state is based on the GLISSANDO 2 \cite{generator, gliss2}  package. This package works in the frame of the semiclassical Glauber model. Within this approach at the very initial stage of the collision, individual interactions between nucleons deposit transverse energy.  Each deposition of the transverse energy at a certain space-time point or region is called a source and each source has its weight which is called relative deposited strength, or RDS. The normalization of RDS can be treated as an insertion of the additional parameter. We choose its value in such a way that all charged-particle multiplicity distribution, obtained in iHKM for 5\% most central collisions, fits its experimental value. The RDS may be different for the wounded nucleons and the binary collisions, and it can fluctuate from source to source. We use a mixed model, amending the wounded nucleon model with some binary collisions. In this model, a wounded nucleon obtains the RDS of $(1-\alpha)/2$ and a binary collision has the RDS of $\alpha$. The total RDS, averaged over events, is then $(1-\alpha)N_W/2+\alpha N_{bin}$. Of course, the result of simulations fluctuates from event to event. In this paper we do not study fluctuations of
observables, and will simulate inclusive observables basing on the approximation of the averaged initial state for each centrality class. This approximation implies that for each selected centrality only the one hydro run is performed.
 The simulation of a single event is made in three stages:
\begin{enumerate}
  \item Generation of the positions of nucleons in the two colliding nuclei according to the fluctuating nuclear density distribution. The form of the mean distribution depends on the mass of nuclei $A$. For sufficiently large nuclei this distribution has a Woods-Saxon form with an account of nuclear deformations, the latter are small in our case of Pb+Pb collisions.
 \item Generation of the transverse positions of the sources and their RDSs.
  \item Calculation of the physical quantities and writing the results in the output file.
\end{enumerate}
In order to obtain the initial conditions from the RDS, one can put the RDS to be proportional to energy or entropy. We choose it to be proportional to energy because entropy is not created yet in a non-equilibrium initial state. Then, for the initial state averaged over many events with a chosen centrality bin, one can write ($b$ is the impact parameter associated with centrality)
\begin{equation}
\epsilon(b;\tau_0, {\bf r}_T) = \epsilon_0 \frac{(1-\alpha)N_W({b,\bf r}_T)/2+\alpha N_{bin}(b,{\bf r}_T)}{(1-\alpha)N_W({b=0,\bf r}_T=0)/2+\alpha N_{bin}(b=0,{\bf r}_T=0)},
\label{RDS}
\end{equation}
where $\epsilon_0$ is a fitting parameter defining the maximal energy density of the initial state that is reached at minimal centrality, say 0--5\% ($b\approx 0$), at ${\bf r}_T=0$. The value $\epsilon_0$ and $\alpha$ are the same for all centrality classes. The values of $\epsilon_0$ and $\alpha$ are fixed by the  reproduction of the charged-particle multiplicity at different centralities.

\subsection{Pre-thermal stage of the matter evolution}
The generation of the initial energy density configuration with the help of GLISSANDO is supposed to be followed by a thermalization process.  As we have already discussed in the Introduction we describe the matter evolution at the pre-thermal stage with the relaxation model \cite{preth, Naboka}. We assume a longitudinal boost invariance, which is a good approximation for the central rapidity region of the fireball at high collision energies. The relaxation model simulations are applied to the system for the early stage of the heavy ion collision, from the proper time $\tau=\tau_0$ to $\tau=\tau_{th}$, where $\tau=\sqrt{t^2-z^2}$. As mentioned above, we use $\tau_0=0.1$ fm/c as the basic value and $\tau_0=0.5$ fm/c for a comparison. Thermalization time is supposed to be $\tau_{th}=1.0$ fm/c, the``conventional'' time of the formation of strongly interaction matter, $t\approx 1/m_{\pi}.$\footnote{Note, that calculations with $\tau_{th}=1.5$ fm/c and with the other parameters to be the same as in our basic scenario (see later) lead to very similar results as calculations with $\tau_{th}=1.0$ fm/c. } The total energy-momentum tensor of the matter is taken in the form \cite{preth, Naboka}
\begin{eqnarray}
T^{\mu \nu}(x)=T^{\mu \nu}_{\text{free}}(x){\cal
P}(\tau)+T_{\text{hydro}}^{\mu \nu}(x)[1-{\cal P}(\tau)], \label{1}
\end{eqnarray}
where $T^{\mu \nu}_{\text{hydro}}(x)$ and $T^{\mu \nu}_{\text{free}}(x)$ are hydrodynamic (local equilibrium) and free (fully or almost free evolving)  components of the energy-momentum tensor, $\cal P(\tau)$ is the weight function satisfying the following conditions: $0 \leq {\cal P}(\tau) \leq 1$, ${\cal P}(\tau_0)=1$, ${\cal P}(\tau_{th})=0$, and $\partial_{\mu}{\cal P}(\tau)_{\tau_{\text{th}}}=0$. Its form will be discussed below. As for the free (or almost free) evolving component, it is well defined in such models as IP-Glasma and EPOS, but not in MC Glauber. To analyze different kinds of anisotropy of the initial state we present the boost-invariant distribution function at the initial hypersurface $\sigma_0$: $\tau=\tau_0$ in the factorized form
\begin{equation}
f(t_{\sigma_0},\textbf{r}_{\sigma_0},\textbf{p}) = \epsilon(b;\tau_0, {\bf r}_T)f_0(p)
\label{f0}
\end{equation}
where $\epsilon(b;\tau_0, {\bf r}_T)$ is defined by Eq. (\ref{RDS}).   At other times $f(x,p)$ is defined by the free-streaming requirement $f(t,\textbf{r},\textbf{p})=f(t_{\sigma_0},{\textbf{r}}-\frac{{\textbf{p}}}{p_{0}} (t-t_{\sigma_0}),\textbf{p})$, so that at all times the free evolution of the energy-momentum tensor is defined by the formula
\begin{eqnarray}
T^{\mu \nu}_{\text{free}}(x)=\int d^{3}p \frac{p^{\mu}p^{\nu}}{p_{0}}f(x,p). \label{2}
\end{eqnarray}
Then the evolution of $T^{\mu \nu}_{\text{free}}(x)$ and corresponding distribution function $f(x,p)$ are defined by the initial conditions, which are generated by GLISSANDO, and the initial momentum anisotropy of the function $f_0(p)$. For ${\cal P}(\tau_0)=1$, we have $T^{\mu \nu}_{\text{total}}(\tau_0,x)=T^{\mu \nu}_{\text{free}}(\tau_0,x)$.

 The hydrodynamic component of the energy-momentum tensor has the form
\begin{eqnarray}
T^{\mu \nu}_{\text{hydro}}(x)= [\epsilon_{\text{hydro}}(x) +
p_{\text{hydro}}(x)+\Pi]u^{\mu}_{\text{hydro}}(x)u^{\nu}_{\text{hydro}}(x)
- [p_{\text{hydro}}(x)+\Pi]g^{\mu \nu} +\pi^{\mu \nu},
 \label{6}
\end{eqnarray}
where $\epsilon_{hydro}$ is the energy density in the local rest frame, $p_{hydro}$ is the pressure, $\pi^{\mu \nu}$ is the shear stress tensor, $\Pi$ is the bulk pressure, and $u^{\mu}_{\text{hydro}}(x)$ is the energy flow four-vector. We neglect the bulk pressure, $\Pi=0$. In curvilinear (hyperbolic) coordinates the equation of motion for the shear stress tensor is taken in the form
\begin{eqnarray}
\langle u^\gamma \partial_{;\gamma} \pi^{\mu\nu}\rangle
=-\frac{\pi^{\mu\nu}-\pi_\text{NS}^{\mu\nu}}{\tau_\pi}-\frac 4 3
\pi^{\mu\nu}\partial_{;\gamma}u^\gamma, \label{evolution-1}
\end{eqnarray}
where the semicolon means the covariant derivation, brackets are defined as $\langle A^{\mu\nu}\rangle=(\frac 1 2 \Delta^\mu_\alpha
\Delta^\nu_\beta+\frac 1 2 \Delta^\nu_\alpha \Delta^\mu_\beta -
\frac 1 3 \Delta^{\mu\nu}\Delta_{\alpha\beta})A^{\alpha\beta}$, $\Delta^{\mu\nu}=g^{\mu\nu}-u^\mu u^\nu$, and $\pi_{NS}$ is the Navier-Stokes shear stress tensor:
\begin{eqnarray}
\pi^{\mu\nu}_\text{NS}&=\eta(\Delta^{\mu\lambda}\partial_{;\lambda}u^\nu+\Delta^{\nu\lambda}\partial_{;\lambda}u^\mu)-\frac
2 3 \eta\Delta^{\mu\nu}\partial_{;\lambda}u^\lambda.
\label{evolution-2}
\end{eqnarray}
For $\cal P(\tau)$ we use the ansatz \cite{preth, Naboka}
\begin{eqnarray}
{\cal P}(\tau)=  \left (
\frac{\tau_{\text{th}}-\tau}{\tau_{\text{th}}-\tau_0}\right
)^{\frac{\tau_{\text{th}}-\tau_0}
 {\tau_{\text{rel}} }},
 \label{3}
\end{eqnarray}
where, as mentioned above, $\tau_0=0.1$ fm/c, $\tau_{th}=1.0$ fm/c, and $\tau_{rel} \leq \tau_{th}-\tau_0$. The value of $\tau_{rel}$ is the model parameter, which characterizes the rate of thermalisation, that is, the speed of conversion of the non-equilibrium state to about the equilibrated one. Writing down the conservation laws for the total energy-momentum tensor in the form $\partial_{;\mu}T^{\mu\nu}_{total}=0$ and accounting that for the free streaming $\partial_{;\mu}T^{\mu\nu}_{free}=0$, we have
\begin{eqnarray}
\partial_{;\mu}\{[1-{\cal P}(\tau)]T^{\mu
\nu}_{\text{hydro}}(x)\}= - T^{\mu
\nu}_{\text{free}}(x)\partial_{;\mu}{\cal P}(\tau). \label{7}
\end{eqnarray}
Let us introduce the re-scaled hydrodynamic tensor $\widetilde{T}^{\mu \nu}_{\text{hydro}}(x)$ with initial conditions $\widetilde{T}^{\mu \nu}_{\text{hydro}}(x)=0$ at $\tau = \tau_{0}$ for all $x$. Then
\begin{eqnarray}
\partial_{;\mu}\widetilde{T}^{\mu
\nu}_{\text{hydro}}(x)= - T^{\mu \nu}_{\text{free}}(x)\partial_{;\mu}
{\cal P}(\tau). \label{8}
\end{eqnarray}
This is the hydrodynamic-type equation of motion with a source. As mentioned above, $T^{\mu \nu}_{free}$ is defined by the the initial state at proper time $\tau_0$, and ${\cal P}(\tau)$ is defined explicitly, then the source in the right-hand side of Eq. (\ref{8}) can be calculated for all $\tau$ and $x$. The introduction of the re-scaled hydrodynamic tensor leads to the rescaling of the shear stress tensor  $\widetilde{\pi}^{\mu\nu} =\pi^{\mu\nu}(1-{\cal P})$. Then, multiplying Eq. (\ref{evolution-1}) by $(1-{\cal P})$, we have the equation of motion for the re-scaled shear stress tensor
\begin{eqnarray}
[1-{\cal P}(\tau)]\left \langle u^\gamma \partial_{;\gamma}
\frac{\widetilde{\pi}^{\mu\nu}}{(1-{\cal P}(\tau))}\right \rangle
=-\frac{\widetilde{\pi}^{\mu\nu}-[1-{\cal
P}(\tau)]\pi_\text{NS}^{\mu\nu}}{\tau_\pi}-\frac {4}{3}
\widetilde{\pi}^{\mu\nu}\partial_{;\gamma}u^\gamma.
\label{evolution-3}
\end{eqnarray}

We also have to specify the equation of state in order to close the set of evolutionary equations. In all calculations the Laine-Schroeder equation of state (EOS) \cite{Laine} is applied.
At $\tau = \tau_{th}$ we switch to the viscous hydrodynamic approach. The equations of motion are nearly the same as in Eq. (\ref{8}), but with the right-hand side equal to zero.

\subsection{Matter evolution in thermal and chemical locally near-equilibrated zone}
At $\tau=\tau_{th}=1$ fm/c the ${\cal P}(\tau_{th})=0$ and the target function is reached: $T^{\mu \nu}(x) = T^{\mu\nu}_{hydro}(x)$. The further evolution is described by the relativistic viscous hydrodynamics according to the equations  (\ref{6}) - (\ref{evolution-2}), (\ref{8}), and (\ref{evolution-3}) with ${\cal P}(\tau)\equiv 0$. A numerical solution of the viscous hydrodynamic equations is constructed with the \texttt{vHLLE} code \cite{vhlle}. Such an evolution describes the expansion of superdense quark-gluon and hadron matter close to local chemical and thermal equilibria with a baryon chemical potential $\mu_B=0$ (which is a good approximation for LHC energies) until the temperature when such an approach breaks down. Then the system has lost the properties of local equilibrium, thermal and chemical as well, and another approximation should be used.

\subsection{Particlization stage}

As discussed in the Introduction, the basic HKM describes particlization as  a continuous process (for review see \cite{hydrokin-3}): particles gradually ``escape'' from the expanding fluid, forming the non-equilibrium Wigner function, which then can be used at some space-like hypersurface as the input for UrQMD hadronic cascade  \cite{hydrokin-2a}. As it was already discussed, at smooth initial conditions the results are very similar to a sudden particlization scenario and we will use the latter in this paper. We assume that the chemically and thermally locally equilibrated evolution takes place until temperature $T=165$ MeV (corresponding to an energy density $\epsilon=0.5$ GeV/fm$^3$ for the Laine-Schroeder EOS) is reached, and switch to particle cascade at the hypersurface defined by this criterion. Such a switching surface is built during the hydrodynamic evolution with the help of the Cornelius \cite{convert} routine. We apply the Cooper-Frye formula to convert the fluid to the cascade of particles:
\begin{eqnarray}
p^0 \frac{d^3 N_i(x)}{d^3 p}=d\sigma_{\mu}p^{\mu}f(p\cdot u(x),T(x),\mu_i(x))
\label{Cooper-Frye1}
\end{eqnarray}
In order to take into account the viscous corrections to the distribution function, Grad's 14-moment ansatz is used. We assume that the corrections are the same for all hadron species. Then formula (\ref{Cooper-Frye1}) transforms to
\begin{eqnarray}
\frac{d^3\Delta N_i}{dp^{*}d(cos\theta)d\phi}=\frac{\Delta \sigma^{*}_{\mu}p^{*\mu}}{p^{*0}}p^{*2}f_{eq}\left(p^{*0};T,\mu_i\right)\left[1+(1\mp f_{eq})\frac{p^*_\mu p^*_\nu \pi^{*\mu \nu}}{2T^2(\epsilon+p)}\right]
\label{Cooper-Frye2}
\end{eqnarray}
This distribution function is used to create an ensemble of particles on the hypersurface. At first, the average number of hadrons of every sort is calculated:
\begin{equation}
\Delta N_i = \Delta \sigma_{\mu}u^{\mu}n_{i,th}=\Delta \sigma ^*_0 n_{i,th}
\label{av1}
\end{equation}
The average total number of particles then equals $\langle N_{tot} \rangle=\Sigma_i N_i$. The exact total number of the particles to be created is sampled according to Poisson distribution with a mean value $<N_{tot}>$. The type of each generated particle is chosen randomly based on probabilities $N_i/N_{tot}$. Then the momentum is assigned to each particle in the local rest frame of the fluid. The direction of momentum is chosen randomly in the $4\pi$ solid angle and its modulus is generated according to the isotropic part of Eq. (\ref{Cooper-Frye2}). After that, the corrections for $W_{residual}\cdot W_{visc}$ are applied via the rejection sampling. The particle position is set to be equal to the position of centroid of the surface element, and the spacetime rapidity is sampled randomly within the longitudinal size of the volume element. Finally, the particle momentum is Lorentz boosted to the center-of-mass frame of the fireball.

\subsection{Hadronic cascade}
The generated hadrons are then fed into the UrQMD cascade. Since the cascade accepts only a list of particles at an equal Cartesian time as an input, the created particles are propagated backward in time to when the first particle was created. The particles are not allowed to interact in the cascade until their trajectories cross the particlization hypersurface. The Laine-Schroeder EoS, that is applied in our analysis corresponds to an equilibrium hadron-resonance gas consisting of about 360 hadrons in the low temperature limit. Many of those heavy hadrons are not included in the UrQMD hadron list. To prevent violation of the energy-momentum conservation even at space-like parts of the isotherm $T=165$ MeV, we decay the heavy resonances, which are not in the UrQMD list, just at the switching hypersurface. The particle propagation is stopped at Cartesian time 400 fm/c, where their coordinates and momenta are recorded. We generate 50000 GLISSANDO events for each centrality class to produce averaged initial energy density profile for the pre-thermal/hydrodynamic evolution. Only the one hydrodynamic calculation is performed for each centrality class with a given set of parameters. For each hydrodynamic evolution 20000 UrQMD events are generated. The generated sets of events are stored in \texttt{ROOT} trees and files, which are further processed with scripts to plot momentum distributions and calculate flow coefficients or correlation functions for physical analysis.

\section{Results and discussion}

\begin{figure}[H]
     \centering
     \includegraphics[width=0.55\textwidth]{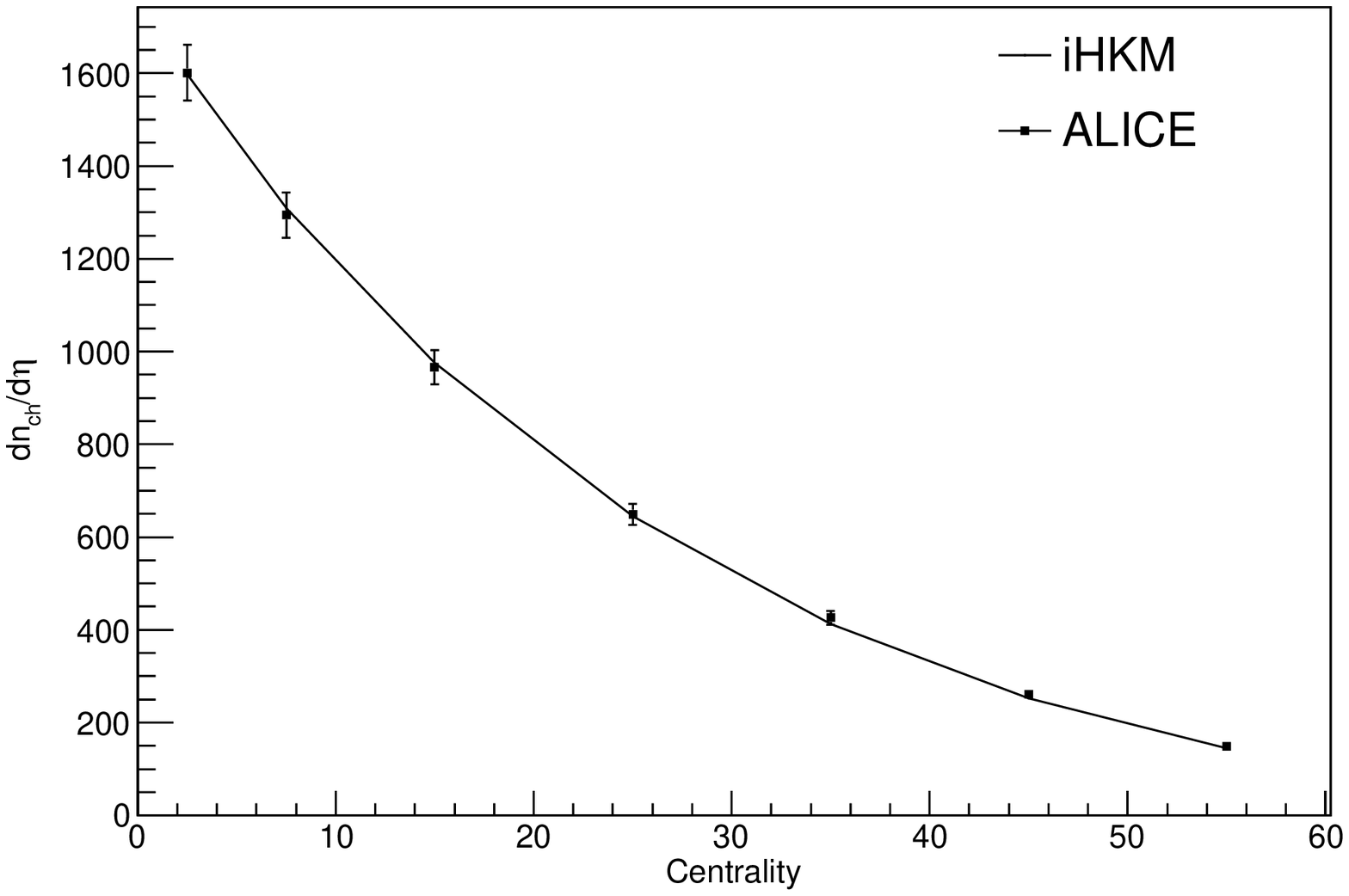}
     \caption{\scriptsize
Multiplicity centrality dependence for the iHKM basic scenario with the parameter values $\tau_0=0.1$ fm/c, $\tau_{rel}=0.25$ fm/c, $\eta/s=0.08$, and $\Lambda=100$ for the LHC energy $\sqrt{s_{NN}}=2.76$ TeV. Experimental data are from \cite{Alica_mult}.}
\label{fig:mult}
\end{figure}

To perform calculations of the spectra, anisotropic flow coefficients, and interferometry radii we need to specify the initial states and model parameters. As discussed above, we generate these initial conditions with GLISSANDO 2, a Monte Carlo Glauber generator. The output of this generator is the relative deposited strength (RDS), which defines the initial energy density (\ref{RDS}) through  the number of wounded nucleons $N_W$ and number of binary collisions $N_{bin}$. The relative weight of these quantities $\alpha$ is crucial for the description of all charged-hadron multiplicities at different centralities. It is found that the value $\alpha = 0.24$ at different reasonable values of other parameters  gives the  best fit for the multiplicities at LHC energy $\sqrt{s}=2.76$ TeV. Figure \ref{fig:mult} demonstrates a typical fit at this value of $\alpha$.

To analyze the effects of momentum anisotropy of the initial state, similarly to Ref. \cite{Naboka} we choose the momentum dependence of the distribution function $f_0(p)$ in Eq. (\ref{f0}) in the form \cite{Rub}:
\begin{eqnarray}
f_0(p)=g \exp\left(-\sqrt{\frac{(p\cdot U)^2-(p\cdot
V)^2}{\lambda_{\perp}^2}+\frac{(p\cdot
V)^2}{\lambda_{\parallel}^2}}\right),
\label{anis1}
\end{eqnarray}
where $U^{\mu}=( \cosh\eta, 0, 0, \sinh\eta)$, $V^{\mu}=(\sinh\eta, 0, 0,\cosh\eta)$. Note, that in the rest frame of the fluid element, $\eta=0$, $(p\cdot U)^2-(p\cdot
V)^2=p_{\perp}^2$ and $(p\cdot
V)^2=p_{\parallel}^2$. Then one can see that the parameters $\lambda_{\parallel}^2$ and $\lambda_{\perp}^2$ can be associated with the different temperatures along the beam axis and orthogonal to it correspondingly. The parameter $\Lambda=\lambda_{\perp}/\lambda_{\parallel}$ is the main parameter, that defines  momentum anisotropy of the initial state.  We use two values for it: $\Lambda=1$ (momentum isotropic case) and $\Lambda=100$ (anisotropic case, almost no pressure in the longitudinal direction). We put $\lambda_{\perp}= 1.4$ GeV \cite{Naboka}. At $m^2/\lambda^2_{\perp}<<1$ the last value is just absorbed into a factor $g=g(\lambda_{\perp},\lambda_{\parallel})$ that is the normalization constant in Eq. (\ref{anis1}). This constant is defined from the requirement that the initial energy-momentum tensor zero component, formally associated with the dimensional function $f_0(p)$ only and so calculated according to Eq. (\ref{2}) with substitution $f(x,p) \rightarrow f_0(p)$, is unity:   $T^{00}[f_0(p)]=1$ [see Eqs. (\ref{RDS}) and (\ref{f0})]. As for the anisotropy factor $\Lambda$, it also does not affect the initial  energy density distribution (\ref{RDS}), but reveals itself in the process of the matter evolution at the pre-thermal stage, see Sec. II B.

The generated distribution functions are related to the two initial times  $\tau_0=0.1$ and $\tau_0=0.5$ fm/c for comparison. The thermalization time is chosen to be, as discussed before, $\tau_{th}=1$ fm/c. When other parameters are fixed,  we make calculations for 0--5\%, 5--10\%, 10--20\%, 20--30\%, 30--40\%, 40--50\%, and 50--60\% centralities with the same normalization coefficient $\epsilon_0$ in Eq. (\ref{RDS}) and fixed relative contribution of binary collision $\alpha = 0.24$. The latter, as we discussed above, does not in fact depend, on all the other parameters, contrary to normalization $\epsilon_0$. The $\epsilon_0$ values at different scenarios, that are associated with iHKM parameters, are shown in Table 1.

\begin{center}
  \begin{tabular}{| c | c | c | c | c | c | c |}
   \hline
   model & $\Lambda$ & $\tau_{rel}$ & $\eta/S$ & $\tau_0$ & $\langle \frac{\chi^2}{ndf} \rangle$ & $\epsilon_0$ [GeV/fm$^3$] \\ \hline
   hydro & - & - & 0 & 0.1 & 5.16 & 1076.5 \\  \hline
   hydro & - & - & 0.08 & 0.1 & 6.93 & 738.8 \\  \hline
   iHKM & 1 & 0.25 & 0.08 & 0.1 & 3.35 & 799.5 \\  \hline
   iHKM & 100 & 0.25 & 0.08 & 0.1 & 3.68 & 678.8 \\ \hline
   iHKM & 100 & 0.75 & 0.08 & 0.1 & 3.52 & 616.5 \\ \hline
   iHKM & 100 & 0.25 & 0.2 & 0.1 & 6.61 & 596.9 \\ \hline
   iHKM & 100 & 0.25 & 0.08 & 0.5 & 5.36 & 126.7 \\ \hline
	\end{tabular}
\end{center}
	Table 1. The maximal initial energy densities $\epsilon_0$ and the mean $\chi^2/ndf$ for different scenarios. The values $\tau_0$, $\tau_{rel}$ correspond to fm/$c$. The mean $\chi^2/$ndf is calculated by an averaging $\chi^2/$ndf values taken from model comparison with observed pion, soft pion, kaon, and antiproton spectra, mid-rapidity charged multiplicity dependence on centrality, $v_2$ transverse momentum dependence for charged particles, and the pion interferometry radii as functions of the transverse momentum.

For each set of parameters with corresponding $\epsilon_0$ (Table 1),  the experimental multiplicity dependence on centrality is reproduced well, similar to Fig. \ref{fig:mult}, with the same $\alpha = 0.24$. There is a  correspondence between the ratios of $\epsilon_0$ at the different parameters and inverse  ratios of the energy densities at the thermalization time $\tau_{th}=1$ fm/c, obtained at the such parameters in \cite{Naboka} when the initial energy densities at $\tau_0$ are the same.

After specifying the initial conditions we run the relaxation model, which describes the pre-thermal stage with different relaxation times $\tau_{rel}=0.25$ and 0.75 fm/c.  The grid spacing and time step for the relaxation model and further pure hydrodynamic calculations are $\delta x = \delta y = 0.2$ fm/c and $\delta \tau = 0.05$ fm/c. Here we perform 2+1-dimensional longitudinally boost-invariant calculations in iHKM, which is a good approximation for the central rapidity interval at LHC energies. At the  thermalization (proper) time  $\tau_{th}=1.0$~fm/c, the evolutionary equations at the pre-thermal stage are smoothly switched (automatically) to relativistic viscous hydrodynamic equations in the Israel-Stewart framework.

We use the two values of the shear viscosity to entropy ratio in the hydrodynamic phase: the minimal one $\eta/s=0.08\approx \frac{1}{4\pi}$ and $\eta/s=0.2$ for comparison. Also we compare the basic scenario (BS) with $\tau_0=0.1$, $\tau_{rel}=0.25$ fm/c, $\eta/s=0.08$, anisotropy parameter $\Lambda=100$, which is selected as giving the optimal description of the experimental data (see below),  with results of calculations with other different parameters, including  viscous and ideal pure hydrodynamic scenarios (without the pre-thermal stage but with the subsequent hadronic cascade).

\begin{figure}
     \centering
     \includegraphics[width=0.8\textwidth]{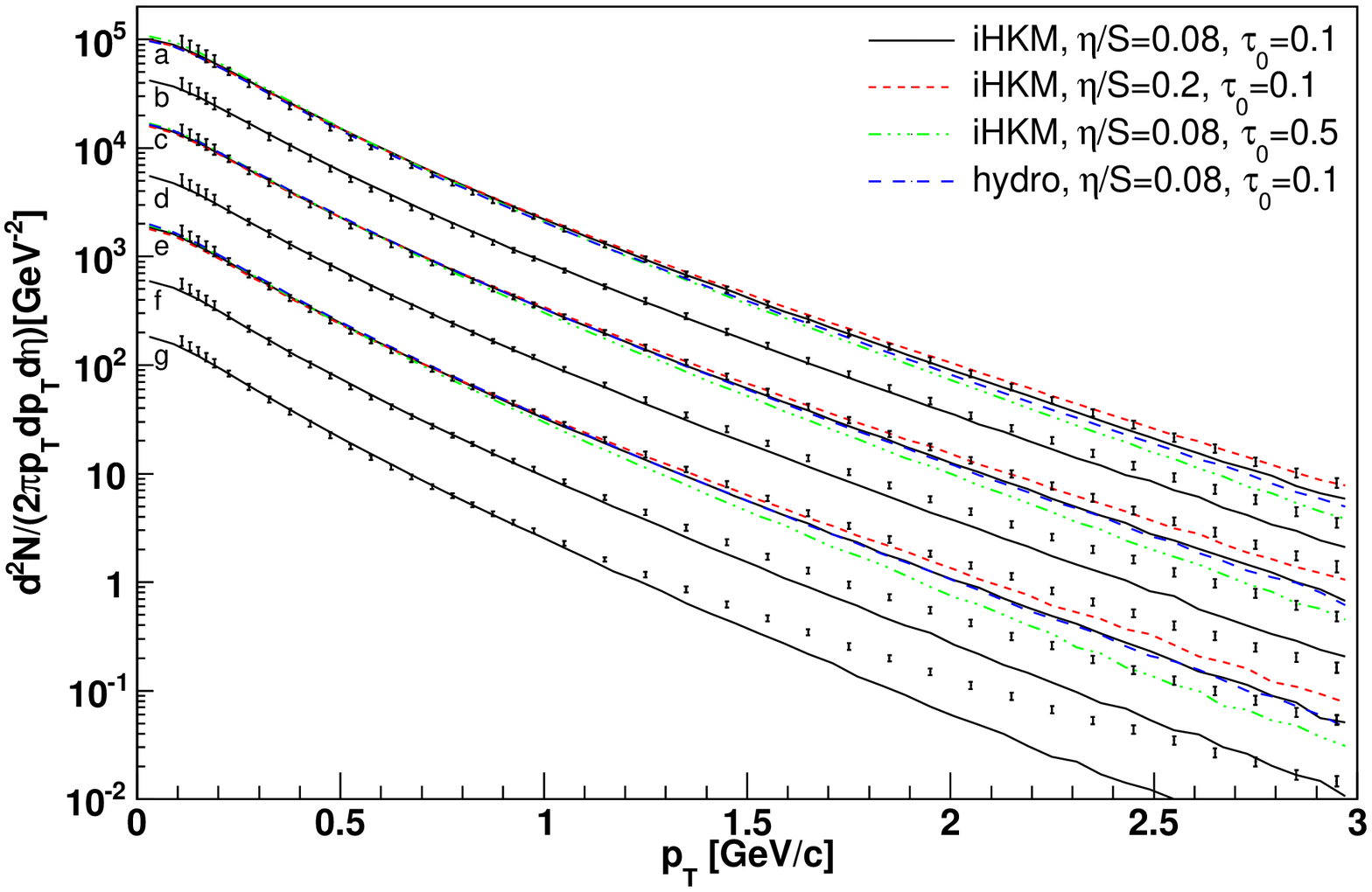}
     \caption{\scriptsize
Resulting pion spectra in the $0.1 <p_T<3$ GeV/\emph{c} region for various centrality classes obtained in the iHKM basic scenario (as in Fig. \ref{fig:mult}). The results are compared with those in iHKM at the other parameter $\tau_0=0.5$ fm/c and with  pure viscous hydro at the starting time $\tau_{th}\rightarrow\tau_0=0.1$ fm/c for centrality classes 0--5\%, 10--20\%, and 30--40\%. The experimental data are from \cite{Alice_spectra}. The spectra for different centralities are multiplied by a factor of 2 ($2^6=64$ for 0--5\% centrality). The lines correspond to different centrality classes as marked by letters: a, 0--5\%; b, 5--10\%; c, 10--20\%; d, 20--30\%, e, 30--40\%; f, 40--50\%; g, 50--60\%}
\label{fig:pi1}
\end{figure}

\begin{figure}
     \centering
     \includegraphics[width=0.8\textwidth]{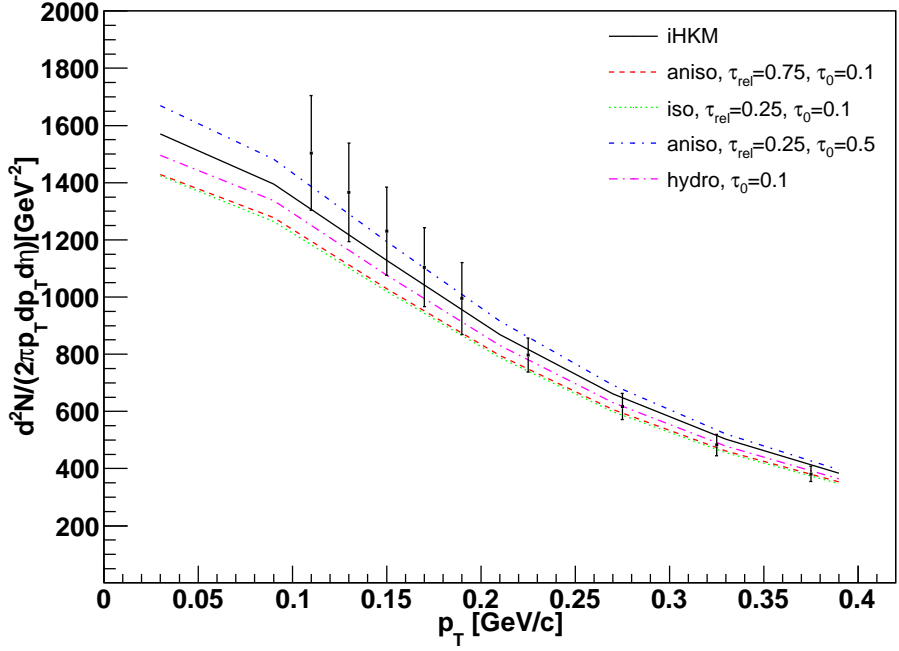}
     \caption{\footnotesize
 Detailed picture of  pion spectra in soft $p_T$ region for 0--5\% centrality in the iHKM basic scenario (as in Fig. \ref{fig:mult}) in comparison with the results obtained with (1) the other relaxation time $\tau_{rel}=0.75$ fm/c instead of 0.25 fm/c, or with (2) the isotropic parameter  $\Lambda=1$ instead of the anisotropy one $\Lambda=100$,  or with (3) the other initial time $\tau_0=0.5$ fm/c instead of 0.1 fm/c. Also the results for pure viscous hydro, starting at $\tau_{th}\rightarrow\tau_0=0.1$ fm/c are presented. The experimental data are from \cite{Alice_spectra}.}
\label{fig:pi2}
\end{figure}

\begin{figure}
     \centering
     \includegraphics[width=0.85\textwidth]{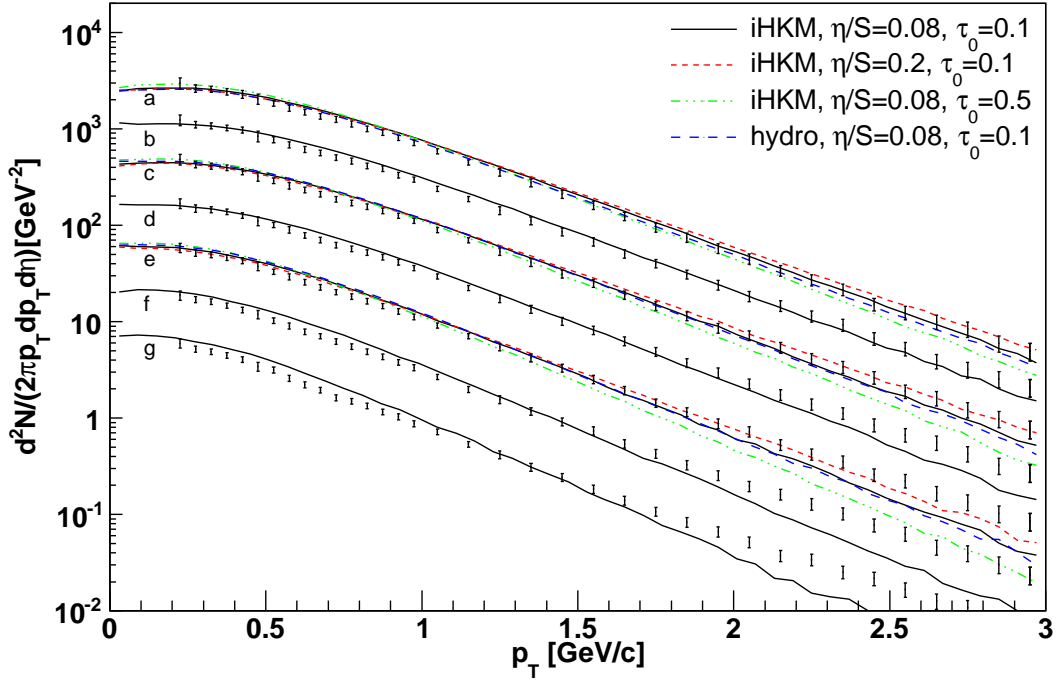}
     \caption{\footnotesize
The resulting kaon spectra under the same conditions as in Fig. \ref{fig:pi1}. }
\label{fig:k1}
\end{figure}

\begin{figure}
     \centering
     \includegraphics[width=0.85\textwidth]{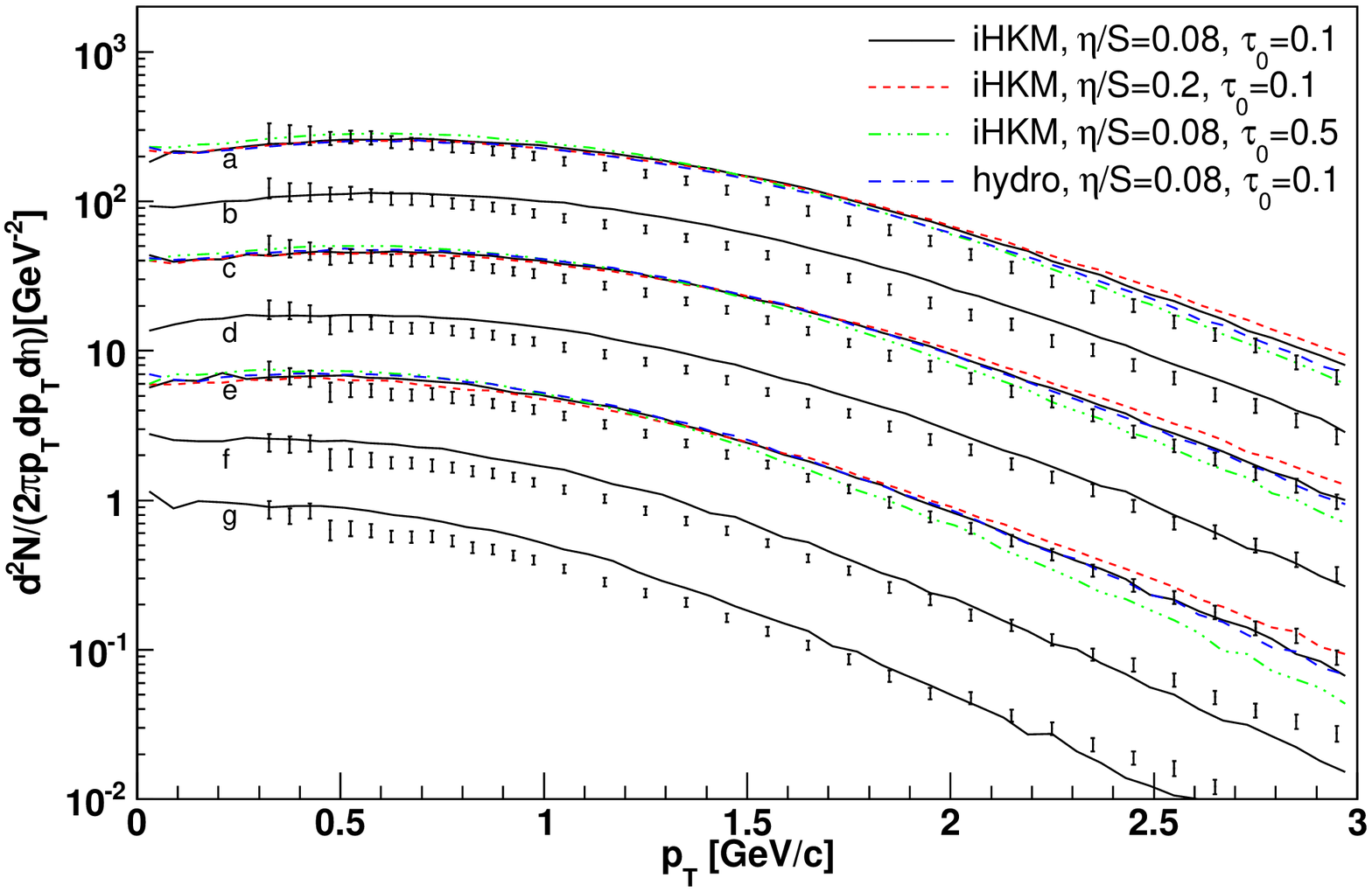}
     \caption{\footnotesize
The resulting antiproton spectra under the same conditions as in Fig.
 \ref{fig:pi1}. }
\label{fig:aproton1}
\end{figure}

\begin{figure}
     \centering
     \includegraphics[width=0.75\textwidth]{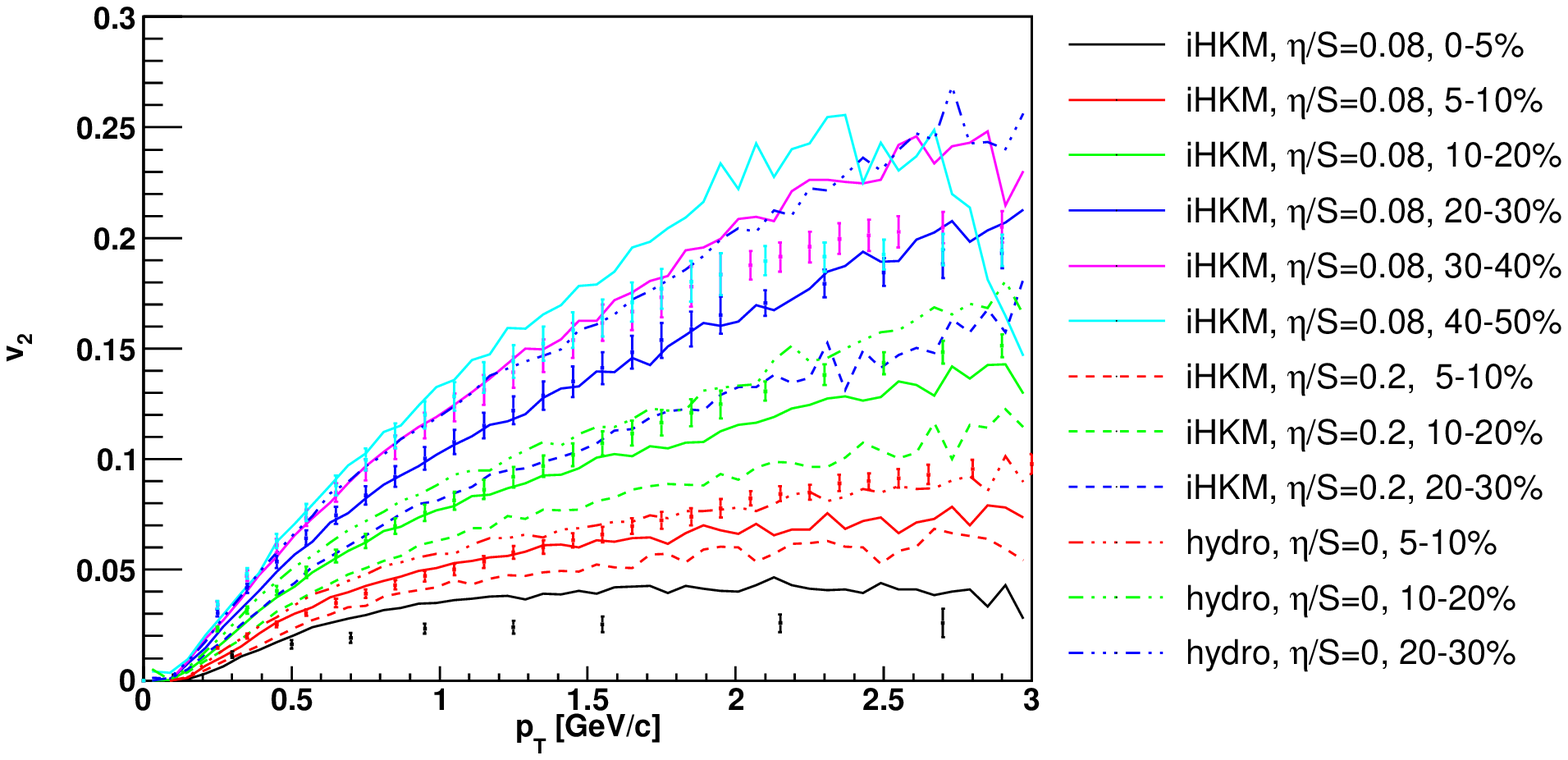}
     \caption{\scriptsize
  All charged-particle $v_2$ coefficients for centrality classes 0--5\%, 5--10\%, 10--20\%, 20--30\%, 30--40\%, 40--50\% and 50--60\%, obtained in the iHKM basic scenario (as in Fig. \ref{fig:mult}). The results are compared with those in iHKM at the other parameter dissipation condition,  $\eta/s=0.2$ instead of $0.08$ and with  ideal hydro with the starting time $\tau_{th}=\tau_0=0.1$ fm/c for centrality classes 5--10\%, 10--20\%, and 20--30\%. The experimental data are from \cite{Alice_v2}.}
\label{fig:v2}
\end{figure}

In Fig. \ref{fig:pi1}  we compare the results with the pion spectra at $0.1<p_T<3$ GeV/\emph{c}.  As one can see at fairly large $p_T > 1.5$ GeV/\emph{c} the results for pion transverse spectra in the scenario with initial time $\tau_0= 0.5$ fm/c describe poorly the experimental data: $\chi^2_{pion}$/ndf=14.33. The best plots correspond to shear viscosity value $\eta/s=0.2$, $\chi^2_{pion}$/ndf=2.48, in the basic scenario (BS) with $\eta/s=0.08 \rightarrow 0.2$. The BS itself leads to satisfactory results for pion spectra also. According to Table 1 the basic scenario leads to mean value $\langle \frac{\chi^2}{\tt{ndf}}\rangle$=3.68 over all  observables under consideration. while the scenario with $\eta/s=0.08 \rightarrow 0.2$ has $\langle \frac{\chi^2}{\tt{ndf}} \rangle$= 6.61 because it does not describe $v_2$ (see below Fig. \ref{fig:v2}), $\chi^2_{v_2}$/ndf=27.92.

In addition to the basic scenario the relatively good description of the wide class of observables gives, according to Table 1, the similar scenarios with, however, larger relaxation time, $\tau_{rel}=0.25$ fm/c $\rightarrow 0.75$ fm/c, and the scenario with isotropic initial state, $\Lambda = 100 \rightarrow 1$. So, for more detailed analysis, let us consider the interesting phenomenon of soft pion enhancement presented in Fig. \ref{fig:pi2}. As one can see, the iHKM scenarios with isotropic initial state, $\Lambda=1$ or with larger relaxation time, $\tau_{rel}=0.75$, obviously underestimate  $p_T$ spectra in the soft region, namely,  $\chi^2_{\tt {soft pions}}$/ndf=1.03 for the isotropic case and $\chi^2_{\tt {soft pions}}$/ndf=1.27 for $\tau_{rel}=$ 0.75 fm/c; pure (viscous) hydrodynamics scenario with $\tau_{th}=\tau_0= 0.1$ fm/c is also less effective at soft $p_T$ for pions. At the same time,  the basic scenario leads to twice less value $\chi^2_{\tt {soft pions}}$/ndf=0.59. The description of this soft momentum region becomes the best when the initial state formation is ascribed to the later time $\tau_0=0.5$ fm/c. However, it excludes, as we discussed above (see also Table 1), a satisfactory spectra description at $p_T>1.5$ GeV/\emph{c} despite that there is no problem with $v_2$ coefficients and the interferometry radii.  This could mean the necessity of including mini-jets in the hydrokinetic picture in order to  describe the large $p_T$ spectra  simultaneously with very small transverse momenta. One of the other proposals considers the specific mechanism of soft pion radiation, namely, the Bose-Einstein condensation in the thermal model \cite{Begun}. However, since the pion spectra in the basic scenario are within the experimental errors with small $\chi^2_{\tt {soft pions}}$/ndf , we consider now the BS as the most realistic, allowing to describe well the bulk observables in wide $p_T$ region.

As one can see from Fig. \ref{fig:k1}, iHKM also describes well the kaon spectra in the basic scenario, $\chi^2_{kaon}$/ndf=1.9.  As for the antiproton spectra (Fig. \ref{fig:aproton1}), one can see that the basic scenario is not the best one (Fig. \ref{fig:aproton1}). That is why, if one compares the mean values of $\langle \frac{\chi^2}{\tt{ndf}}\rangle$ over all observables in Table 1, this scenario brings  slightly worse results than scenarios with an isotropic initial state and anisotropic one with larger relaxation time. At the same time the last two scenarios do not describe the enhancement of soft pions as we already discussed.

\begin{figure}
     \centering
     \includegraphics[width=0.75\textwidth]{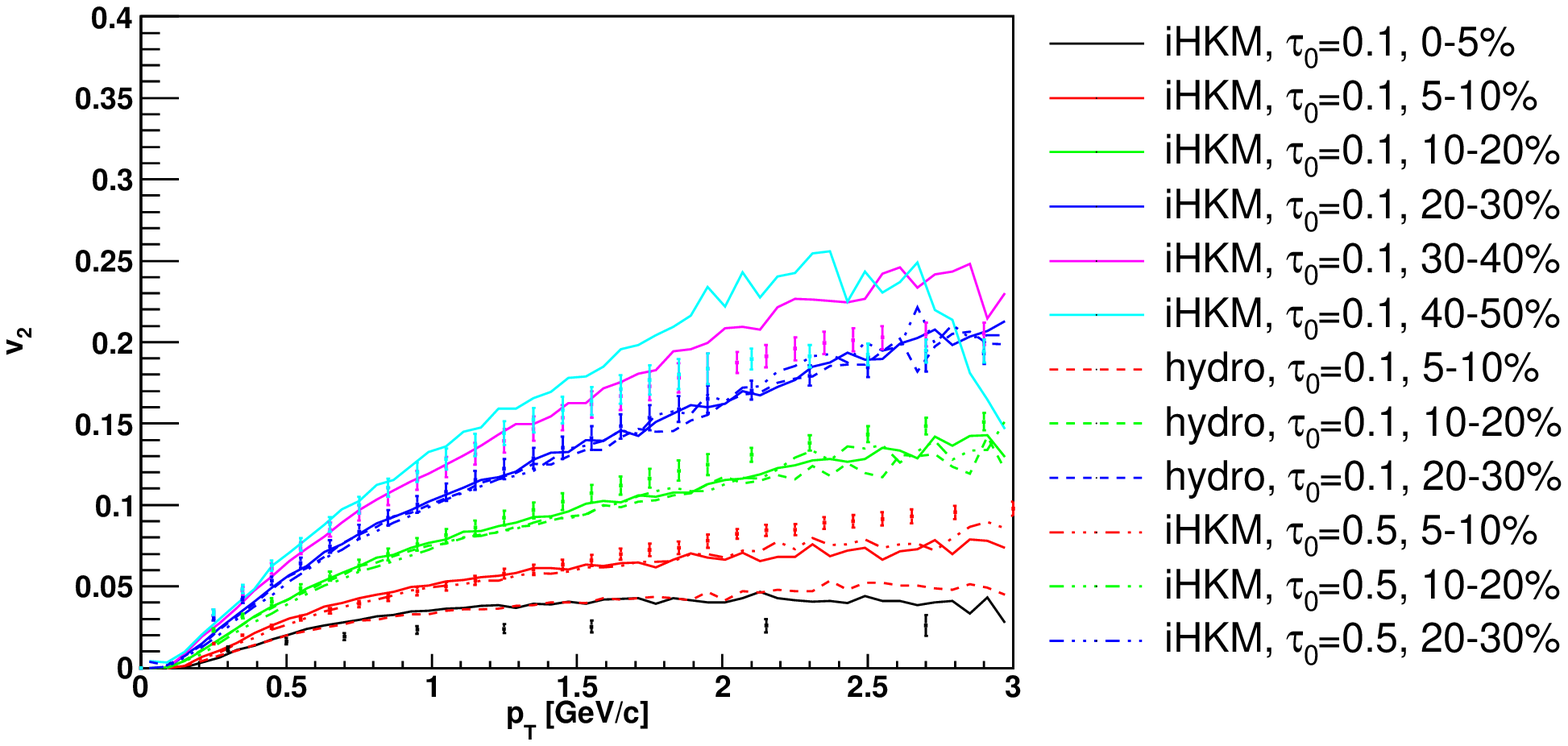}
     \caption{\scriptsize
 Same as Fig. \ref{fig:v2}, but the results are compared with those in iHKM at the other initial time,  $\tau_0=0.5$ fm/c instead of $0.1$ fm/c and with  viscous hydro at the starting time $\tau_{th}=\tau_0=0.1$ fm/c.}
\label{fig:v2_2}
\end{figure}

Let us apply iHKM to a description of elliptic flows.  The elliptic flows in iHKM are associated with
the reaction plane that is well-defined in GLISSANDO 2. We compare the results with the experimental $v_2$-coefficients   estimated
with four-particle cumulants measured for unidentified charged particles as a function of
transverse momentum for various centrality classes, $v_2\{4\}$ \cite{Alice_v2}.
Figure \ref{fig:v2}  demonstrates all charged-particle $v_2$-coefficients for the basic scenario in comparison with iHKM results with $\eta/s=0.2$ instead of $\eta/s=0.08$ and also with ideal hydro starting at
$\tau_{th}=\tau_0= 0.1$ fm/c. One can see that the two last scenarios lead to poorly described $v_2$-coefficients. At the same time the basic scenario, as well as the iHKM one with $\tau_0=0.5$ instead of $\tau_0=0.1$ fm/c and also the viscous pure hydrodynamics scenario with $\tau_{th}=\tau_0= 0.1$ fm/c describe the data quite satisfactorily, see Fig. \ref{fig:v2_2}. However, the  two last scenarios describe the spectra worse: the one with $\tau_0=0.5$, at fairly large $p_T$; the other, the pure viscous hydro, at both large and small momenta.

An important observation for evolutionary models is the interferometry radii that reflect the space-time structure of the particle emission from an expanding fireball. The detailed study within iHKM shows that if a selection of some set of parameter values, such as $\tau_0$, $\tau_{rel}$, $\Lambda$, $\eta/s$,  is accompanied by the renormalization of the maximal initial energy density $\epsilon_0$, similar to that in Table 1, to keep the iHKM charged-particle multiplicity to be  equal to the experimental one, the interferometry radii change within a few percent only, which is a much smaller change than the one caused by a change in the centrality class of the collision. Published results (see, e.g. review \cite{hydrokin-3}) have shown that the interferometry radii depend on particle species, multiplicity, and initial system size. The other details are not so important.

\begin{figure}
     \centering
     \includegraphics[width=0.8\textwidth]{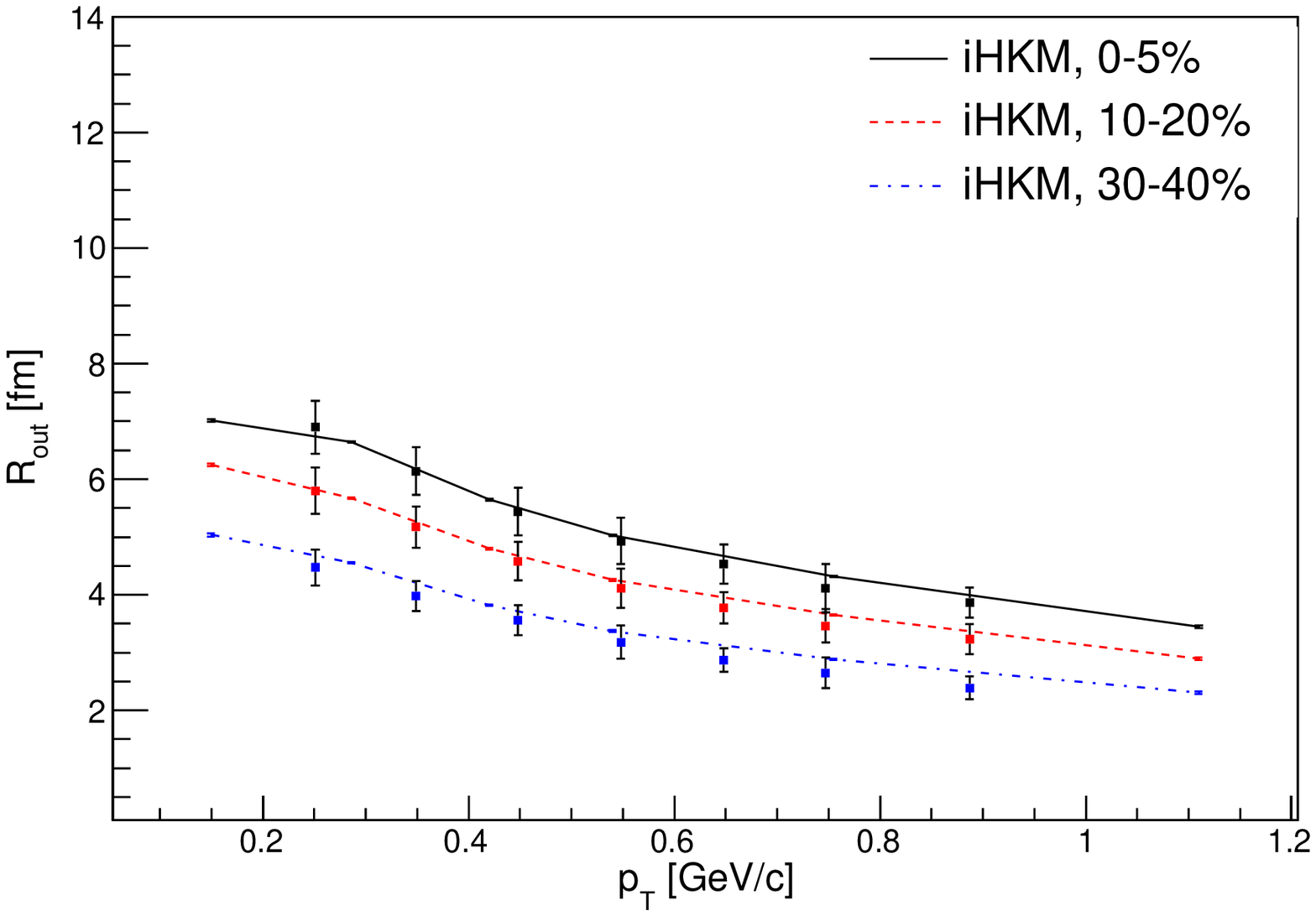}
     \caption{\small (Color online)
The $R_{out}$ dependence on transverse momentum for different centralities in the iHKM basic scenario under the same conditions as in Fig. \ref{fig:mult}.}
\label{fig:Rout}
\end{figure}

\begin{figure}
     \centering
     \includegraphics[width=0.8\textwidth]{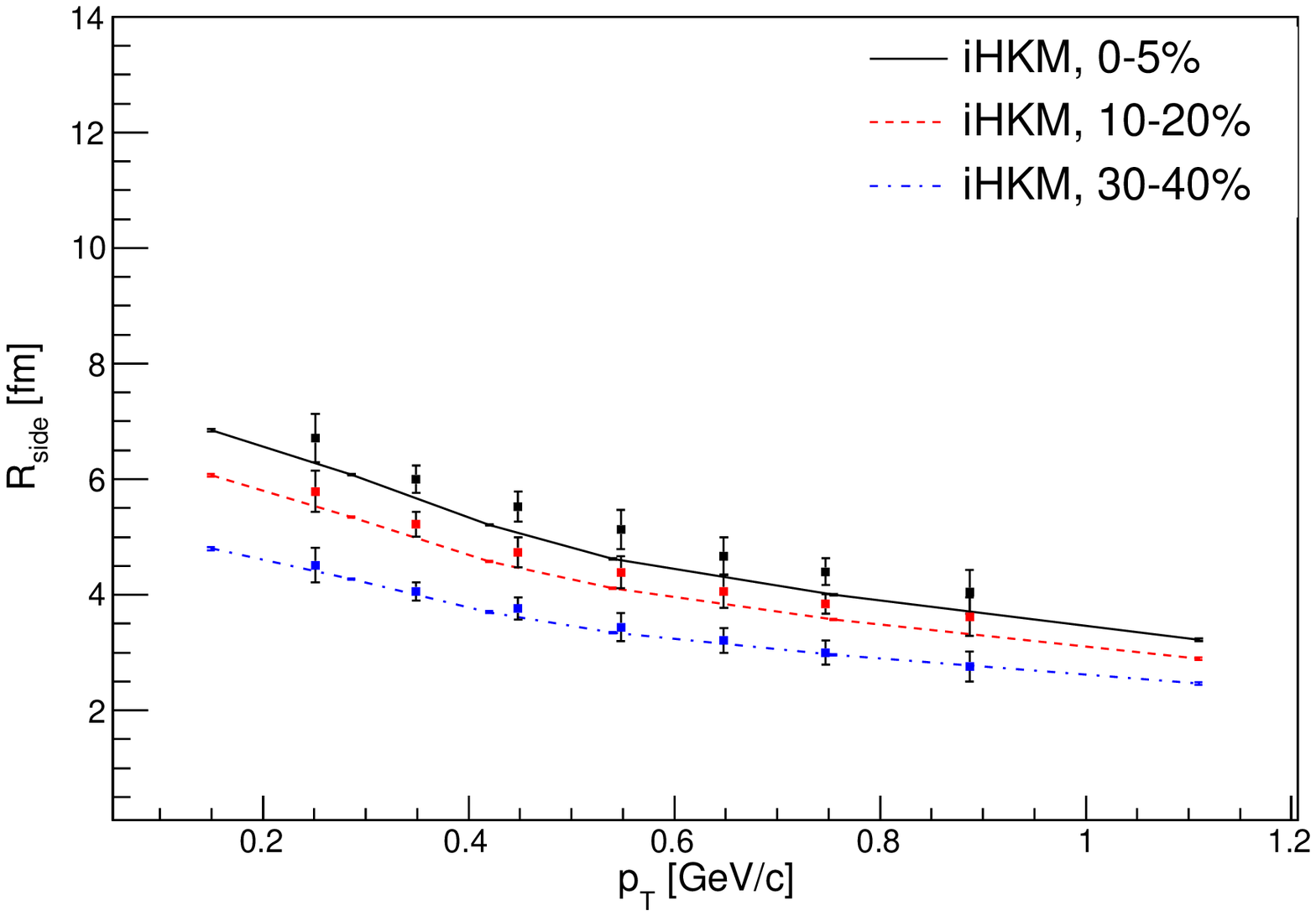}
     \caption{\small (Color online)
The $R_{side}$ dependence on transverse momentum for different centralities in the iHKM basic scenario under the same conditions as in Fig. \ref{fig:mult}. The experimental data are from \cite{Alice_radii}.}
\label{fig:Rside}
\end{figure}

\begin{figure}
     \centering
     \includegraphics[width=0.8\textwidth]{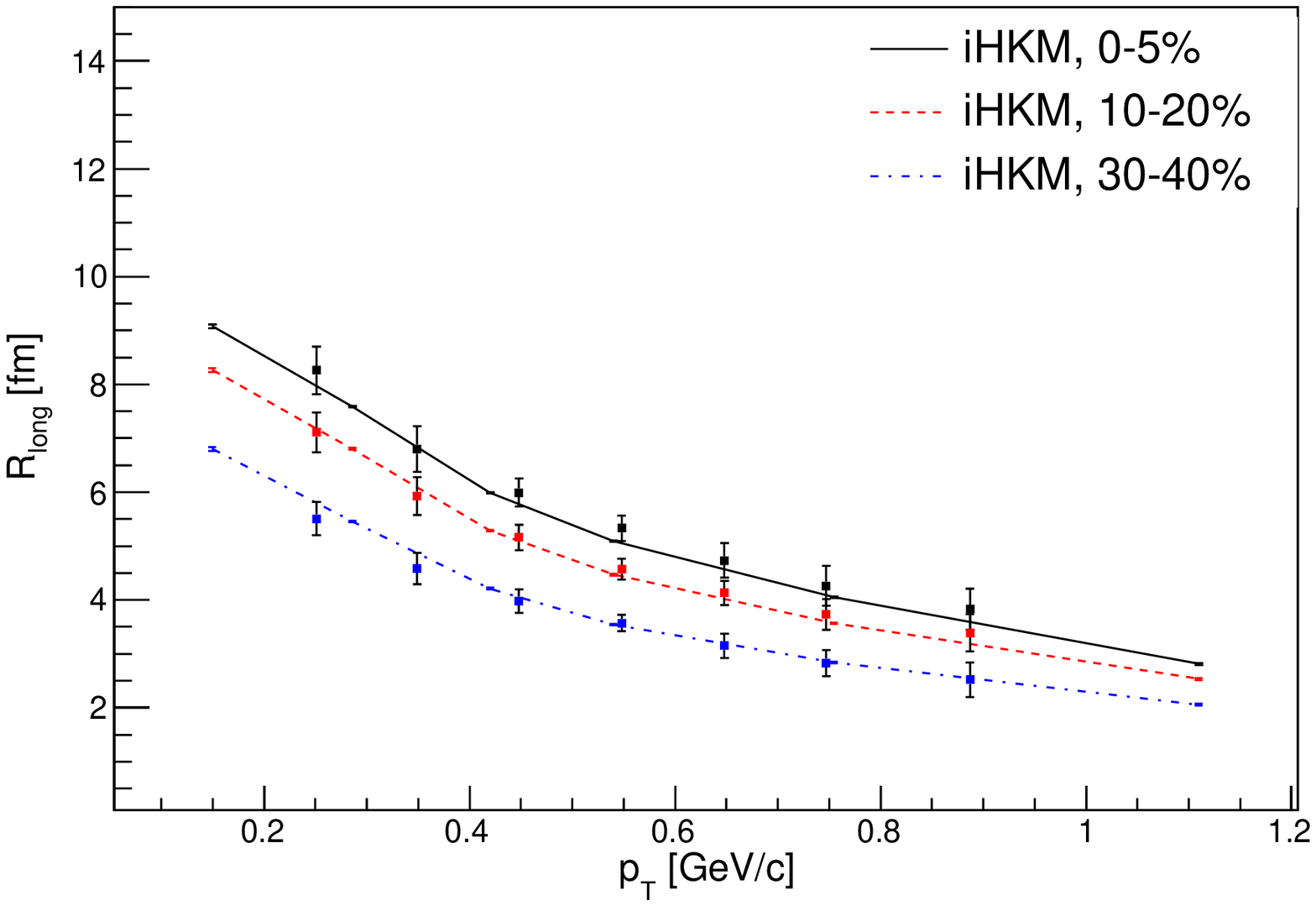}
     \caption{\small (Color online)
The $R_{long}$ dependence on transverse momentum for different centralities in the iHKM basic scenario under the same conditions as in Fig. \ref{fig:mult}.  The experimental data are from \cite{Alice_radii}.}
\label{fig:Rlong}
\end{figure}

\begin{figure}
     \centering
     \includegraphics[width=0.8\textwidth]{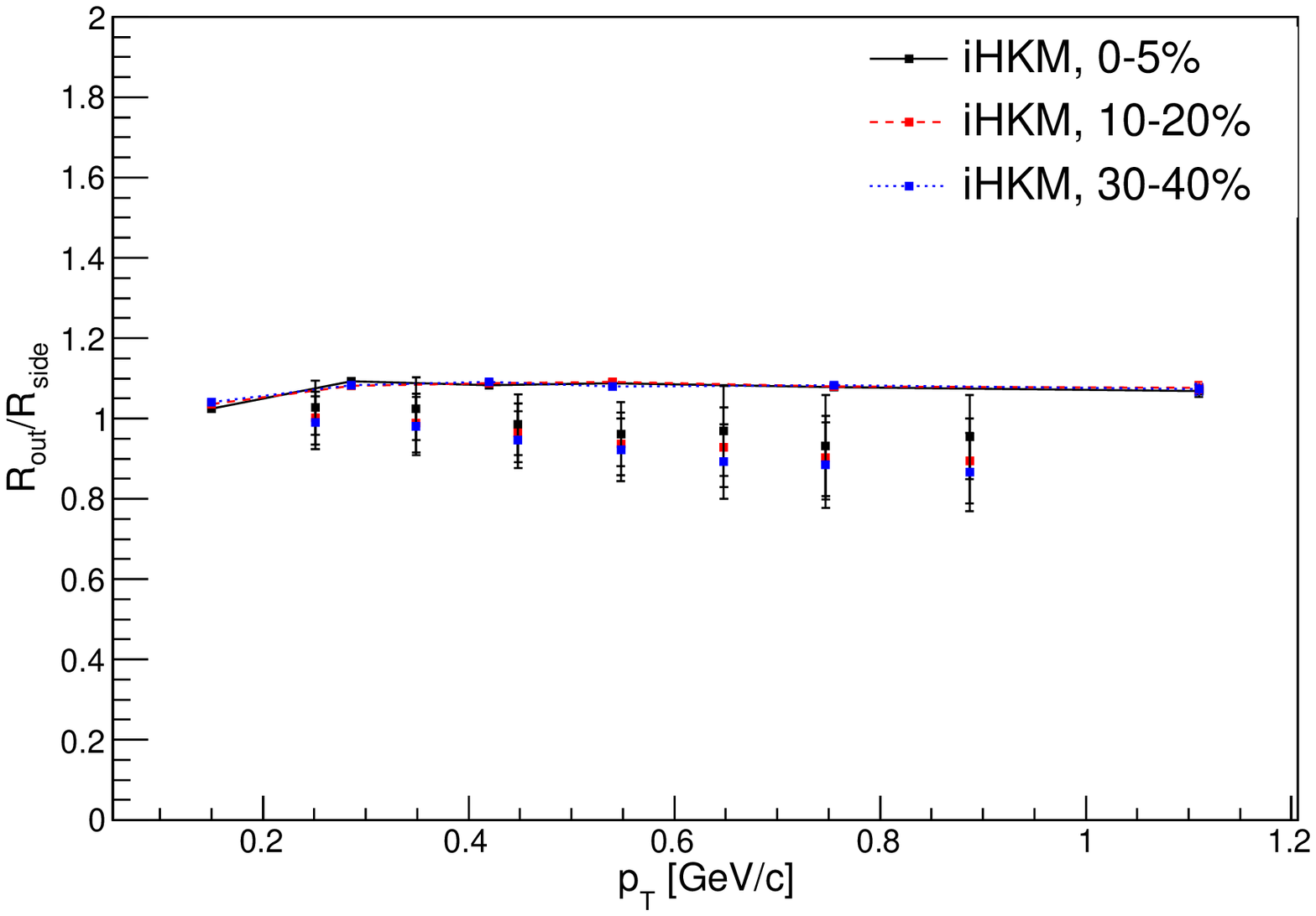}
     \caption{\small (Color online)
The $R_{out}/R_{side}$ ratio for different centralities in the iHKM basic scenario under the same conditions as in Fig. \ref{fig:mult}.  The experimental data are from \cite{Alice_radii}.}
\label{fig:RoRs}
\end{figure}

In Figs. \ref{fig:Rout}--\ref{fig:Rlong} we present the pion interferometry radii in the iHKM  basic scenario for three different centralities. In general the radii are described well except some small deviation for $R_{side}$ and $R_{out}$ from the central experimental points; these deviations are oppositely directed  for these transverse radii, which results in slight overestimation of the $R_{out}$ to $R_{side}$ ratio, as demonstrated in Fig. \ref{fig:RoRs}.

\section{Summary}
The integrated hydrokinetic model (iHKM) of \emph{A+A} collisions is developed. It includes the generation of the initial (generally momentum anisotropic) state, matter evolution at the pre-thermal stage leading to thermalization, subsequent viscous hydrodynamic expansion, particlization, and hadronic cascade UrQMD. This model is applied to describe at different centralities the mid-rapidity charged-particle multiplicities, the pion, kaon, and antiproton spectra, the $v_2$ coefficients for all charged hadrons, and the pion interferometry radii in Pb-Pb collisons at LHC with $\sqrt{s_{NN}}=2.76$ TeV. It is found that a quite satisfactory description of the bulk observables can be reached when the initial states are attributed to the very small initial
time 0.1 fm/c,  the pre-thermal stage (thermalization process)   lasts at least until  1 fm/c, and the shear viscosity at the hydrodynamic stage of the matter evolution has its minimal value, $\eta/s = \frac{1}{4\pi}$. The requirement of satisfactory description of the soft pion enhancement phenomenon also leads to the further discrimination between the iHKM scenarios, namely, the initial state should be maximally momentum anisotropic and have a small mean relaxation time,  $\tau_{rel}\ll \tau_{th}$. In this basic scenario the bulk observables are described in a wide $p_T$ region, including the pion and antiproton spectra, as well as their yields when  annihilation processes at the afterburner (UrQMD) stage are taken into account \cite{hydrokin-2a}.

 It is worthy noting that while the found tendencies, which direction
the changes of spectra and $v_2$ values go when changing the basic model parameters, are probably stable, an accounting of event-by-event fluctuations, as well as modification of such important factors as EoS and particlization temperature, can change the numerical values of the model parameters, ensuring an optimal description of the observables. Such an investigation is the subject of subsequent work.

Another important point which we would like to stress is the approximate similarity of the results at different reasonable values of the main parameters of the model, even including the pure hydrodynamic scenario starting from $\tau_{th}\rightarrow \tau_0=0.1$ fm/c. The reason for such similarity  is probably that each change of basic parameters is accompanied by a re-normalization of the maximal energy density of the initial state, to keep the agreement with the experimental multiplicity of all charged particles at midrapidity. Then, at the same particlization temperature for all the scenarios, the main characteristics of the bulk observables are approximately preserved. Of course, some details are different, as iHKM demonstrates, but the variations are not so dramatic. This observation explains satisfactory agreement of  different variants of hydrodynamic and hybrid models with the experimental data for A+A collisions, especially when not all bulk observables are considered.

In this paper the Monte Carlo Glauber model is used for the generation of the initial state. The profile from GLISSANDO is associated here with the one for energy density of the non-equilibrium  and non-thermal initial state. Note that the entropy density cannot be used to characterize such a state at $\tau_0=0.1$ fm/c. The interesting point is that with such treatment of the Monte Carlo Glauber model, the share of the binary collisions, contributed to the total energy density, is constant at about 1/4 for LHC energies at {\it different}  iHKM parameters. We are planning to utilize other initial state models, such as EPOS and IP-Glasma, which have other structures of the initial state and already fixed values of the maximal initial energy density, as the generators of the initial states in iHKM. The comparison of the models have to be done on an event-by-event basis and include besides spectra, their anisotropy and femtoscopic scales for different hadron pairs, as well as fluctuations of the observables.

\begin{acknowledgments}
The research was carried out within the scope of the EUREA: European Ultra Relativistic Energies Agreement (European Research Group: ``Heavy ions at ultrarelativistic energies''), and is  supported by the National Academy of Sciences of Ukraine (Agreements F7-2016). I.K. acknowledges financial support by the Helmholtz International Center for FAIR and the Hessian LOEWE initiative.

\end{acknowledgments}

\end{document}